\documentclass[conference]{IEEEtran}
\IEEEoverridecommandlockouts
\usepackage{amsmath,amssymb,amsfonts}
\usepackage{algorithmic}
\usepackage{graphicx}
\usepackage{textcomp}
\usepackage{xcolor}
\usepackage{multirow}
\usepackage{booktabs}

\usepackage[style=ieee]{biblatex} 
\def\BibTeX{{\rm B\kern-.05em{\sc i\kern-.025em b}\kern-.08em
    T\kern-.1667em\lower.7ex\hbox{E}\kern-.125emX}}
\begin{document}

\title{Advancing Diagnostic Precision: Leveraging Machine Learning Techniques for Accurate Detection of Covid-19, Pneumonia, and Tuberculosis in Chest X-Ray Images}



\author{
  \IEEEauthorblockN{Aditya Kulkarni\textsuperscript{1}, Guruprasad Parasnis\textsuperscript{1}, Harish Balasubramanian\textsuperscript{1}, Vansh Jain\textsuperscript{2}, Anmol Chokshi\textsuperscript{2}, and Reena Sonkusare\textsuperscript{3}}
  \IEEEauthorblockA{\textsuperscript{1} Department of Electronics and Telecommunication, Sardar Patel Institute of Technology, Mumbai}
  \IEEEauthorblockA{\textsuperscript{2} Department of Information Technology, Sardar Patel Institute of Technology, Mumbai}
  \IEEEauthorblockA{\textsuperscript{3} Head of Department of Electronics and Telecommunication, Sardar Patel Institute of Technology, Mumbai}
}

\maketitle

\begin{abstract}

Lung diseases such as COVID-19, tuberculosis (TB), and pneumonia continue to be serious global health concerns that affect millions of people worldwide. In medical practice, chest X-ray examinations have emerged as the norm for diagnosing diseases, particularly chest infections such as COVID-19. Paramedics and scientists are working intensively to create a reliable and precise approach for early-stage COVID-19 diagnosis in order to save lives. But with a variety of symptoms, medical diagnosis of these disorders poses special difficulties. It is essential to address their identification and timely diagnosis in order to successfully treat and prevent these illnesses. In this research, a multiclass classification approach using state-of-the-art methods for deep learning and image processing is proposed. This method takes into account the robustness and efficiency of the system in order to increase diagnostic precision of chest diseases.  A comparison between a brand-new convolution neural network (CNN) and several transfer learning pre-trained models including VGG19, ResNet, DenseNet, EfficientNet, and InceptionNet is recommended. Publicly available and widely used research datasets like Shenzen, Montogomery, the multiclass Kaggle dataset and the NIH dataset were used to rigorously test the model. Recall, precision, F1-score, and Area Under Curve (AUC) score are used to evaluate and compare the performance of the proposed model. An AUC value of 0.95 for COVID-19, 0.99 for TB, and 0.98 for pneumonia is obtained using the proposed network. Recall and precision ratings of 0.95, 0.98, and 0.97, respectively, likewise met high standards. Using chest X-rays, the proposed model was found to be remarkably accurate at detecting chest diseases in various scenarios and conditions. Our testing data set shows that the suggested model outperforms competitors' approaches significantly, achieving state-of-the-art results, and creating a benchmark for medical diagnosis through chest scans.

\end{abstract}


\section{Introduction}
In recent years, the field of medical imaging has witnessed a remarkable transformation with the advent of machine learning techniques. These methods have demonstrated great potential in enhancing diagnostic accuracy, improving patient care, and aiding healthcare professionals in making critical decisions \cite{1}. Among the numerous applications, the automated detection and classification of respiratory diseases using chest X-ray images have emerged as a significant area of research.
The outbreak of the Covid-19 pandemic has highlighted the pressing need for accurate and rapid diagnostic tools to combat infectious diseases effectively. In addition to Covid-19, pneumonia and tuberculosis continue to pose significant global health challenges, causing millions of deaths each year. Traditional diagnostic approaches for these diseases often rely on human expertise, which can be subjective, time-consuming, and prone to errors. However, the integration of machine learning algorithms with chest X-ray analysis has shown promising results in terms of early detection, efficient screening, and precise classification \cite{2}.
This paper aims to provide a comprehensive review and analysis of machine learning methods utilized for the detection and classification of Covid-19, pneumonia, and tuberculosis using chest X-ray images. By analyzing the latest advancements in this rapidly evolving field, we seek to explore the potential of these intelligent systems in revolutionizing healthcare practices.
The paper will delve into the application of machine learning techniques for automated detection and classification. We will explore various methodologies, including deep learning algorithms, convolutional neural networks (CNNs), ensemble models \cite{3}, and also pre-trained models which have been successfully employed in previous studies. Emphasis will be placed on the pre-processing steps, feature extraction methods, and model architectures employed by these algorithms to achieve accurate disease identification. The paper will introduce a novel CNN method that works better than conventional algorithms and pre-trained models \cite{4}. 
Furthermore, we will discuss how each model can also provide a better approach to predicting and classifying a particular disease. 

\section{Related Work}

Machine Learning algorithms have been implemented in areas of biomedical applications to reduce the time and cost of the disease identification process. The Covid-19 pandemic was one such event that led to multiple researchers focusing on new methodologies to detect Covid-19 instead of using RT-PCR kits. \cite{5} shows the use of deep learning methods including K Nearest Neighbours, Radial Basis Function, Naive Bayes, Support Vector Machine, and Multi-Layer Perceptron on CBC parameters and achieved a testing accuracy of 88.26\% using One R classifier which will still not guarantee if that person has Covid-19 and in such a critical area such a low accuracy cannot be trusted in real life. Through this approach, we saw the potential of predicting Covid-19 using Machine Learning. Another method present to detect Covid 19 is using InceptionNet and DenseNet in \cite{6}. In this method, the use of chest X-rays and CT images are used to detect COVID-19 and the best accuracy achieved is 92.35\% using InceptionNet and X-Ray images. However this accuracy is not applicable to medical applications. We also noted using \cite{7}, \cite{8}, and \cite{9} that Pneumonia can also be detected using the chest X-rays. In \cite{10}, pre-trained CNN models XCeption, VGG-16, VGG-19, ResNet-50, DenseNet-121, and DenseNet-169 were implemented, and ResNet-50 with an SVM classifier provided the best-case AUC of 0.77. \cite{11} derived 0.76 AUC using 121-layered DenseNet trained on Chest X-Ray (named CheXNet). 13 more pathologies were present, and the model could identify them all. \cite{12} also achieves a 0.76 AUC using DenseNet161 and a boosted cascade ConvNet. After further research, it was noted that tuberculosis can be also detected using the Chest X-Ray, and \cite{13} uses the same deep learning approach to achieve a recall of 0.83 and 0.67 precision. \cite{14} uses pre-trained ResNet and EfficientNet to achieve an accuracy of 89\%. A hybrid approach of Computer-Aided Detection and CNN is suggested in \cite{15} which gives a 92.54\% accuracy and \cite{16} uses ensemble learning to get 93.59\% accuracy. Through this paper, we propose a multi-class approach to detect Covid-19, Pneumonia, and Tuberculosis using a proposed CNN method and compare it with pre-trained models of DenseNet, VGG-19, ResNet, EfficientNet, and InceptionNet and improve the accuracy of detection even further so that this method is usable in real-life medical applications. 

\section{Proposed Methodology}

\subsection{Dataset}\label{AA}
The dataset utilized in this study was sourced from Kaggle, and structured into three main folders (train, test, val), each containing subfolders categorizing the X-ray images into Normal, Pneumonia, COVID-19, and Tuberculosis classes. This dataset comprises a total of 7135 X-ray images. Our proposed approach was subjected to evaluation on various datasets dedicated to pneumonia, tuberculosis, and COVID-19 detection. Furthermore, we conducted tests using the benchmark NIH dataset, an extensive repository comprising 112,120 frontal chest X-ray images showcasing a wide range of anomalies. Our aim was to demonstrate the effectiveness of our method by achieving commendable results on this diverse dataset.

\subsection{Data Pre-Processing}
The dataset has been pre-processed to provide better results owing to various transformations and modifications. The paper introduces a data augmentation technique to adjust the brightness and contrast factor in each image for better recognition and understanding capabilities. It was observed that after augmentation of the data, the results obtained were better than those results when the data was used in its original form. 

\begin{figure}[htbp]
  \begin{minipage}[t]{0.26\textwidth}
    \centering
    \includegraphics[width=\textwidth]{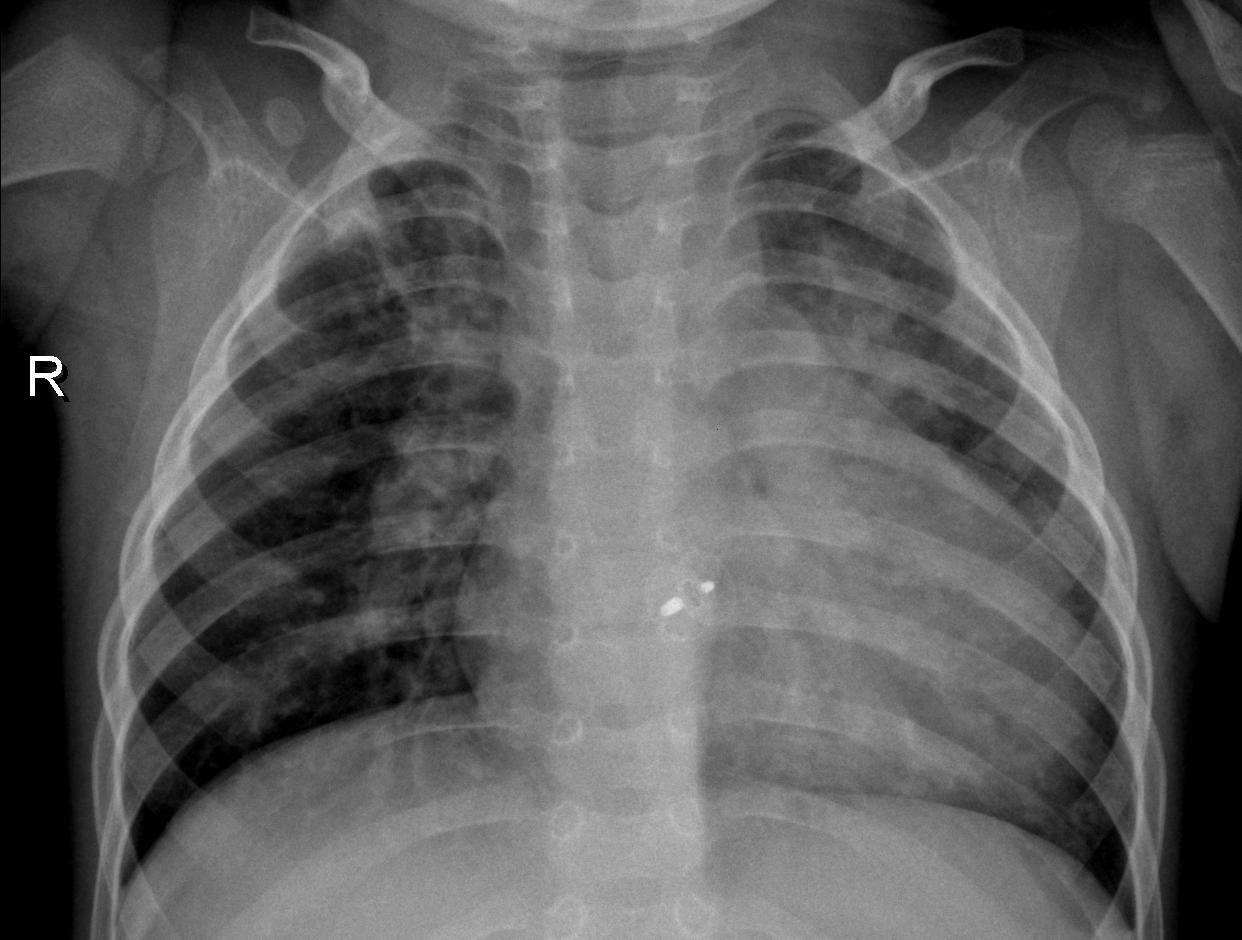}
  \end{minipage}
  \hfill
  \begin{minipage}[t]{0.20\textwidth}
    \centering
    \includegraphics[width=\textwidth]{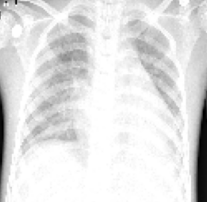}
  \end{minipage}
  \caption{Data Pre-processing by Augmentation}
  \label{fig1:combined}
\end{figure}
 Fig. 1 shows the augmented image which is obtained by enhancing the brightness and contrast. It provides a more detailed textural feature to focus upon than the original image. It is these augmented images that form an integral part of the proposed methodology. \par 
Out of the other image pre-processing techniques that the paper experimented upon, a significant one was the Local Ternary Pattern (LTP) which enhances the texture of the original image better. LTP is a texture descriptor used for image analysis and pattern recognition. It captures the local texture information by encoding the relationship between a central pixel and its neighboring pixels. LTP is particularly useful in scenarios where robustness to noise and illumination variations is required. LTP operates by comparing the intensity value of a central pixel with its surrounding pixels. The comparison is done using three labels: +1 (for pixels greater than or equal to the central pixel value), -1 (for pixels less than the central pixel value), and 0 (for pixels with the same value as the central pixel). Fig. 2 shows how the textures are highlighted after applying LTP on the original images. 

\begin{figure}[htbp]
  \begin{minipage}[t]{0.24\textwidth}
    \centering
    \includegraphics[width=\textwidth]{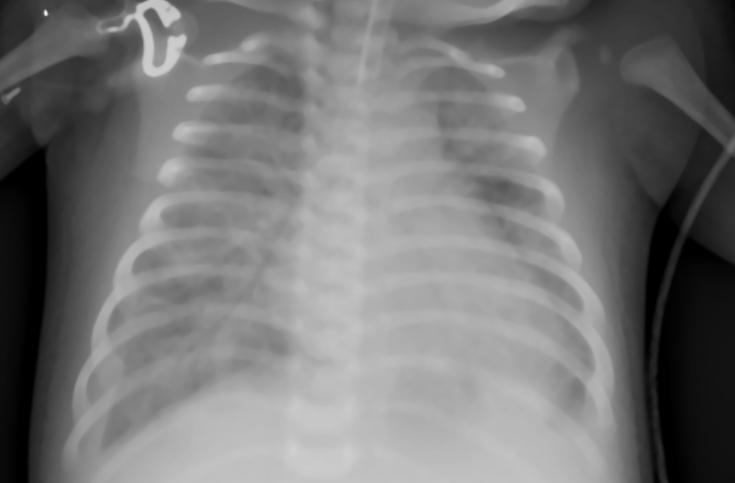}
  \end{minipage}
  \hfill
  \begin{minipage}[t]{0.24\textwidth}
    \centering
    \includegraphics[width=\textwidth]{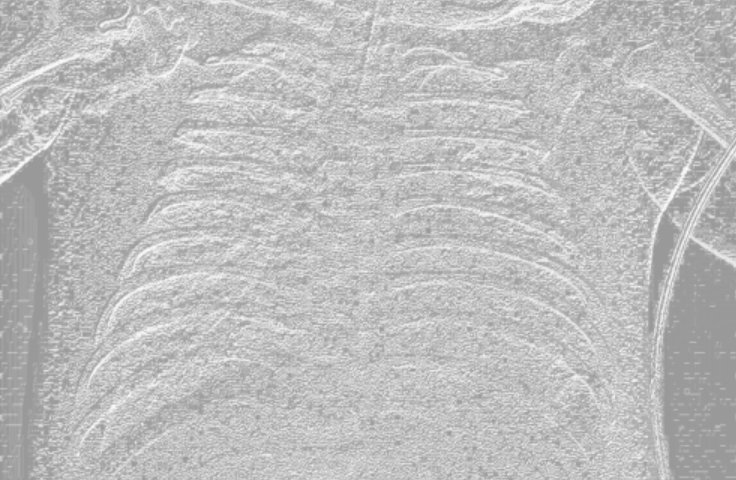}
  \end{minipage}
  \caption{Data Pre-processing by LTP}
  \label{fig2:combined}
\end{figure}
The method also focuses on histogram equalization as a possible image processing technique to enhance the features and textures present in the X-ray scan images. Histogram equalization is a technique used to enhance the contrast of an image by redistributing the pixel intensities. It works by transforming the histogram of the image such that the resulting histogram becomes more uniformly distributed. This technique can be applied to grayscale images. By applying the histogram equalization technique, the resulting image will have a histogram that is approximately flat, which improves the contrast of the image. This is because the transformation redistributes the intensities across a wider range, utilizing the full dynamic range of the image. Fig. 3 shows how the histogram equalization technique enhances the original image. 
\begin{figure}[htbp]
  \begin{minipage}[t]{0.24\textwidth}
    \centering
    \includegraphics[width=\textwidth]{pneu_hist_og.png}
  \end{minipage}
  \hfill
  \begin{minipage}[t]{0.24\textwidth}
    \centering
    \includegraphics[width=\textwidth]{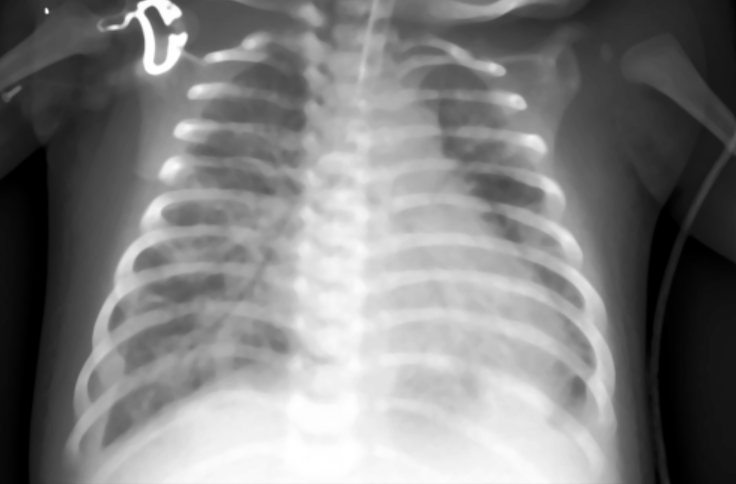}
  \end{minipage}
  \caption{Data Pre-processing by Histogram Equalization}
  \label{fig3:combined}
\end{figure}
The paper also analyses the adaptive thresholding method which is a popular image processing technique for converting grayscale images to binary images by converting pixel values above a certain threshold to white and those below it to black. Adaptive thresholding is a type of thresholding in which the threshold value for each pixel is adjusted based on the local intensity of its surroundings. This technique is useful for our images with varying contrast or non-uniform illumination. Fig. 4 shows how the adaptive thresholding technique transforms the original image. \par
\begin{figure}[htbp]
  \begin{minipage}[t]{0.24\textwidth}
    \centering
    \includegraphics[width=\textwidth]{pneu_hist_og.png}
  \end{minipage}
  \hfill
  \begin{minipage}[t]{0.24\textwidth}
    \centering
    \includegraphics[width=\textwidth]{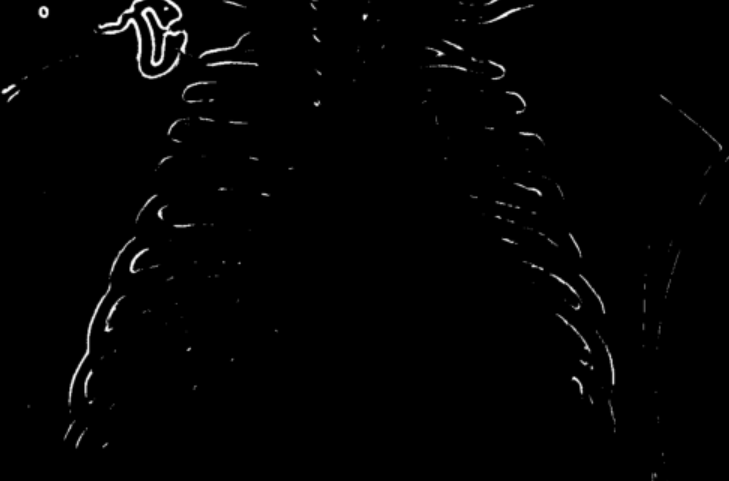}
  \end{minipage}
  \caption{Data Pre-processing by Thresholding}
  \label{fig4:combined}
\end{figure}
By comparing all the above-experimented image processing methods, it is clear that augmentation performs the best, closely followed by histogram equalization. The main difference between these methods is that in the augmentation technique, the infected part is enhanced more than the surrounding part of the bones and rib cages. Hence, while performing classification tasks on the proposed model, the augmented images perform better than the images which have histogram equalization applied to them.

\subsection{Performance Metrics}
The paper focuses on some pre-trained deep learning networks to test the performance in the multiclass classification task of detecting anomalies in chest X-ray scans. The paper uses multiple such models: namely VGG19, InceptionNet, ResNet, EfficientNet and DenseNet. These models are also tested along with a transfer learning block to experiment with the performance metrics of the model. \par 
The performance metrics that the paper deals with are precision, recall, F1-score and AUROC. In the multiclass classification task, the results are obtained by providing values for these metrics for each of the classes in the dataset for each of the networks that are used in this paper. \par 
Precision measures the proportion of correctly predicted positive instances out of all instances predicted as positive. It focuses on the accuracy of positive predictions. \par
\[
\text{Precision} = \frac{\text{True Positives}}{\text{True Positives} + \text{False Positives}}
\]

True Positives (TP): The number of correctly predicted positive instances.\par 
False Positives (FP): The number of negative instances incorrectly predicted as positive. \par 
A high precision score indicates a low number of false positives, meaning that the model is good at identifying true positives and has a low rate of wrongly classifying negatives as positives. \par 

Recall calculates the proportion of correctly predicted positive instances out of all actual positive instances. It focuses on the ability of the model to find all positive instances. \par 
\[
\text{Recall} = \frac{\text{True Positives}}{\text{True Positives} + \text{False Negatives}}
\]
False Negatives (FN): The number of positive instances incorrectly predicted as negative.
A high recall score indicates that the model is good at capturing positive instances, minimizing the number of false negatives. \par 

The F1 score is the harmonic mean of precision and recall. It provides a single metric that balances the trade-off between precision and recall. The F1 score is particularly useful when dealing with imbalanced datasets.
\[
\text{F1 Score} = \frac{2 \cdot \text{Precision} \cdot \text{Recall}}{\text{Precision} + \text{Recall}}
\]

The F1 score ranges between 0 and 1, with 1 indicating the best possible model performance.\par 

The AUC is a metric used to evaluate the performance of binary classification models based on their Receiver Operating Characteristic (ROC) curve. The ROC curve plots the true positive rate (TPR) against the false positive rate (FPR) at different classification thresholds.

\begin{figure*}
\centering
  \includegraphics[width=1\linewidth]{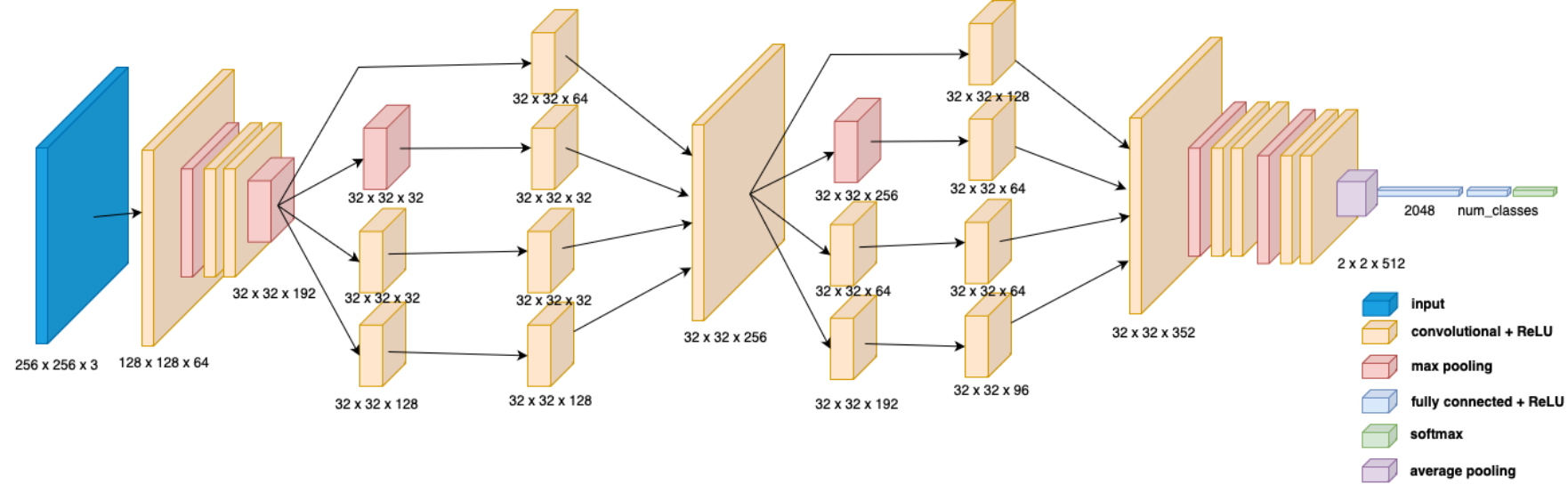}
  \caption{Proposed Deep learning network}
  \label{fig6}
\end{figure*} 

\subsection{Proposed Deep Learning Network}
The paper studies all the pre-trained models, their hyperparameters and adaptations of the network to various situations of data splitting, model layers, etc. Based on the complete analysis, the paper proposes a deep learning architecture that performs better than all of the networks against which it has been compared. \par 
\begin{figure}[htbp]
\centerline{\includegraphics[width=0.4\textwidth]{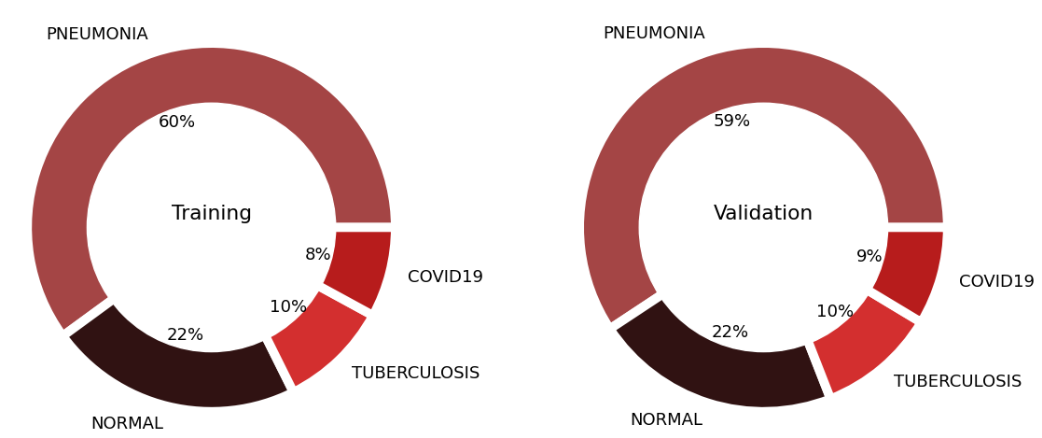}}
\caption{Data Splitting}
\label{fig5}
\end{figure}


The paper introduces a unique approach to processing and splitting the data for effective use in the classification task. As described in the section above, the processing is done with the help of augmentation. For the splitting of the data, the paper defines a new function to split the data according to pre-defined batch size, the processing technique, and the number of epochs. The images in each folder are split accordingly and are fed to different training and validation data generators. This generates random training and testing data by separating them from their individual classes and then combining them into one non-separate training dataset. This splitting makes the classification task more efficient and robust. Fig. 5 shows a visual representation of the splitting of the data into training and validation sets.

Figure 5 demonstrates the proposed network including all the steps and the architecture.

The new network architecture features:

\begin{enumerate}
    \item Initial Convolution Layers: The model begins with a convolutional layer (7x7) followed by a max pooling layer. This is followed by a series of convolution layers leading into the inception-like modules.
    \item Custom Inception-like Module: Instead of traditional stacking of convolutional layers, the architecture employs custom inception modules. Within these modules, there are multiple parallel branches that capture different spatial characteristics. The branches include:
   - A 1x1 convolution.
   - A 3x3 convolution following a 1x1 convolution.
   - A 5x5 convolution following a 1x1 convolution.
   - Max pooling followed by a 1x1 convolution.
   The outputs from these branches are concatenated together. The model incorporates two of these custom inception modules in succession.
   \item Further Convolution Layers: Post inception modules, the model integrates a sequence of convolutional layers with 256 and 512 filters.
   \item Final Layers: After processing through the aforementioned convolutional and inception layers, the model reduces dimensionality using average pooling with a 7x7 filter. This is followed by flattening and a dense layer for classification.
   \item Output Layer: The final dense layer has the number of neurons equal to the number of classes in the dataset, with a softmax activation function to classify the input images.
\end{enumerate}

\section{Experimental Results}
In this paper, we experiment with the various pre-processing methods mentioned to determine the most suitable one for the classification task. The network utilizes preprocessed chest x-ray images with adaptive thresholding, LTP, histogram equalization and augmentation techniques as input images. 
The images after applying thresholding are seen in Fig. 7. 
\begin{figure}[htbp]
\centerline{\includegraphics[width=0.4\textwidth, height = 5cm]{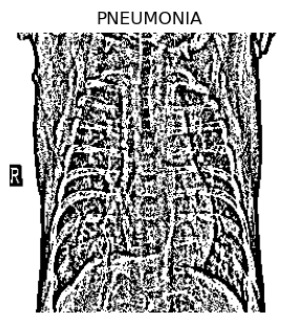}}
\caption{Input image by Adaptive Gaussian Thresholding}
\label{fig7}
\end{figure}
On using the thresholded images in the proposed network, we obtain the following results from Table I. The network performs with a strikingly good AUC value of 0.99 and a precision value of 0.9799 for the tuberculosis class. In contrast, the COVID-19 class performs very poorly, with an AUC value of 0.88 due to the limited feature enhancement done by the thresholding method. The network also performs decently well with an AUC value of 0.93 on the pneumonia class but is not competent enough with the state-of-the-art. 
The thresholding method is also applied tuberculosis dataset too. Table II showcases the results of the proposed approach when thresholding is applied as pre-processing to the input images.

\begin{table}
\centering
\caption{Results of the thresholded images in the proposed Network on the multiclass dataset}
\begin{tabular}{lcccc}
\toprule
\textbf{Disease} & Recall & Precision & F1-Score & AUC \\
\midrule
Covid-19 & 0.7642 & 0.9691 & 0.8545 & 0.88 \\
Normal & 0.8317 & 0.8955 & 0.862 & 0.90 \\
Pneumonia & 0.9716 & 0.9278 & 0.9492 & 0.930 \\
Tuberculosis & 0.9799 & 0.9299 & 0.9542 & 0.99 \\
\bottomrule
\end{tabular}
\end{table}

\begin{figure}[htbp]
\centerline{\includegraphics[width=0.5\textwidth]{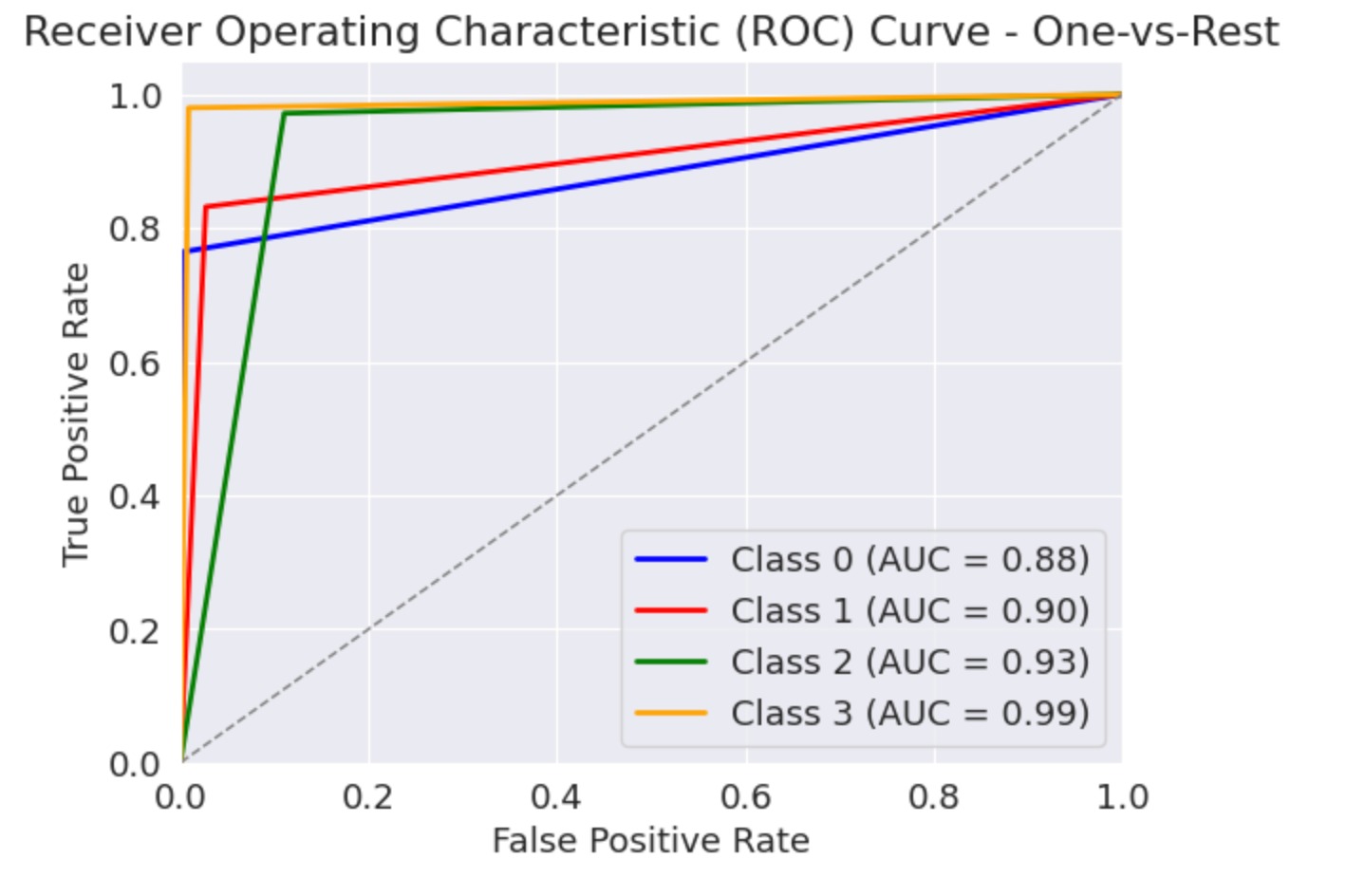}}
\caption{AUC for the multiclass dataset with thresholded images}
\label{fig8}
\end{figure}

\begin{table}
\centering
\caption{Results of the thresholded images in the proposed Network on the tuberculosis dataset}
\begin{tabular}{lcccc}
\toprule
\textbf{Disease} & Recall & Precision & F1-Score & AUC \\
\midrule
Normal & 0.9951 & 0.9951 & 0.9951 & 0.98 \\
Tuberculosis & 0.9750 & 0.9750 & 0.9750 & 0.98 \\
\bottomrule
\end{tabular}
\end{table}

\begin{figure}[htbp]
\centerline{\includegraphics[width=0.5\textwidth]{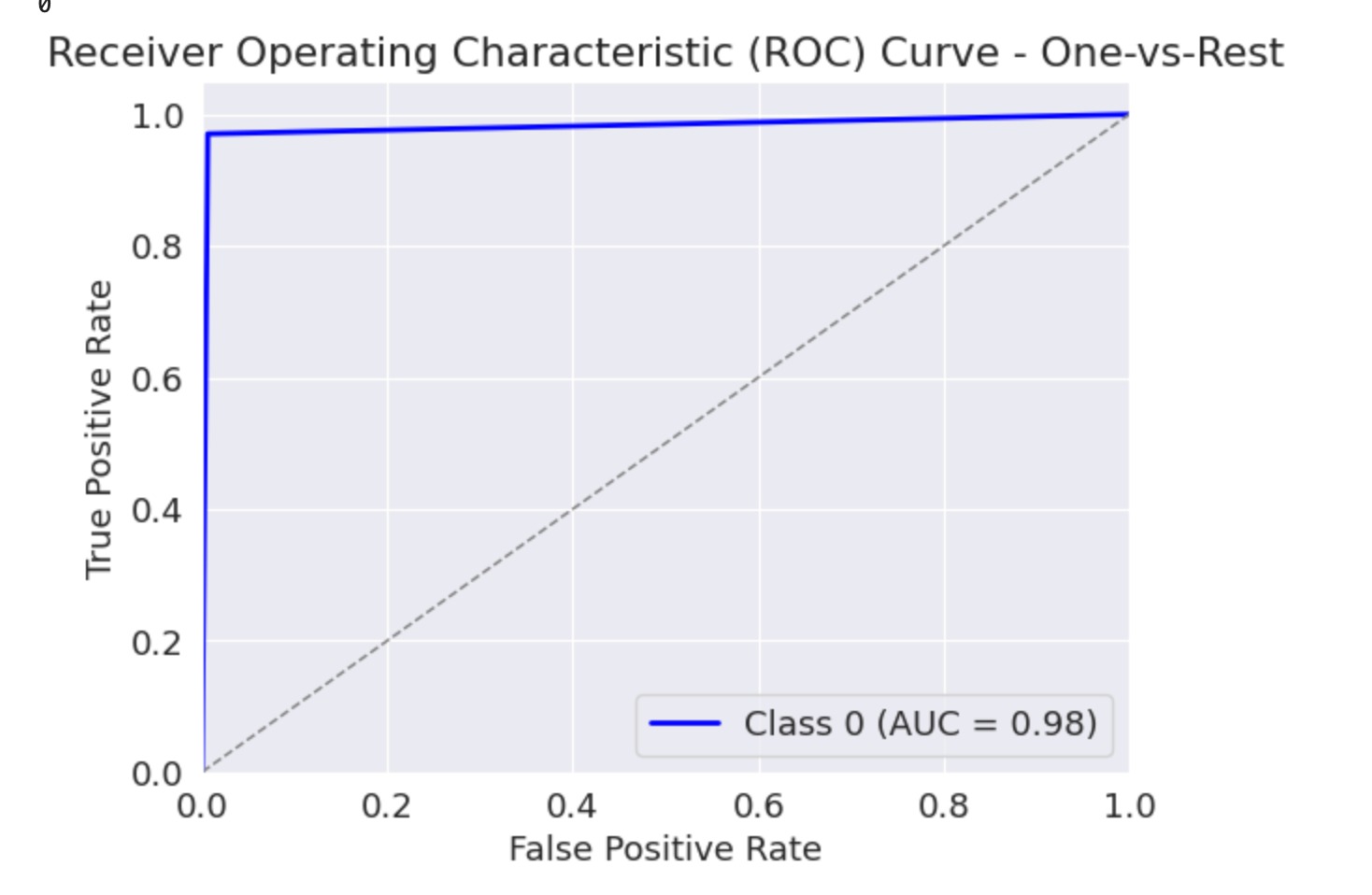}}
\caption{AUC for the tuberculosis dataset with thresholded images}
\label{fig9}
\end{figure}

In the tuberculosis dataset, the thresholded images do perform at a very competent level. This is due to the fact that the dataset contains only two classes and is less diverse. The goal of this paper is to design a network that is robust and efficient in all types of datasets with state-of-the-art performance. Fig. 9 provides the AUC value for the tuberculosis dataset upon the application of thresholded images. 

The paper then experiments with the histogram equalization method for the datasets and evaluates the performance metrics for the same. The chest X-ray images with histogram equalization preprocessing are seen in Fig. 10. 
\begin{figure}[htbp]
\centerline{\includegraphics[width=0.4\textwidth, height = 5cm]{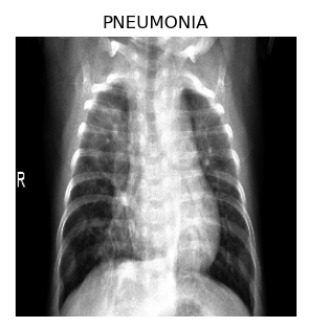}}
\caption{Input images with histogram equalization}
\label{fig10}
\end{figure}
On using the histogram equalization images in the proposed network, we obtain the following results from Table III. The network performs with consistently competent AUC values for all four classes. Moreover, the highly diverse COVID-19 class performance metrics are state-of-the-art too, with an AUC of 0.95 and a recall value of 0.9412. The F1 scores for each of the classes too are state-of-the-art which is an indication of an excellent balance between precision and recall. The F1 score for pneumonia is 0.9732 while for tuberculosis, it is 0.9463, signifying a strong robustness. Fig. 11 shows the AUC values for the multiclass dataset for histogram equalized images. 

\begin{table}
\centering
\caption{Results of the histogram equalized images in the proposed Network on the multiclass dataset}
\begin{tabular}{lcccc}
\toprule
\textbf{Disease} & Recall & Precision & F1-Score & AUC \\
\midrule
Covid-19 & 0.9106 & 0.9412 & 0.9256 & 0.95 \\
Normal & 0.9029 & 0.9654 & 0.9331 & 0.96 \\
Pneumonia & 0.9870 & 0.9598 & 0.9732 & 0.96 \\
Tuberculosis & 0.9463 & 0.9463 & 0.9463 & 0.97 \\
\bottomrule
\end{tabular}
\end{table}

\begin{figure}[htbp]
\centerline{\includegraphics[width=0.5\textwidth]{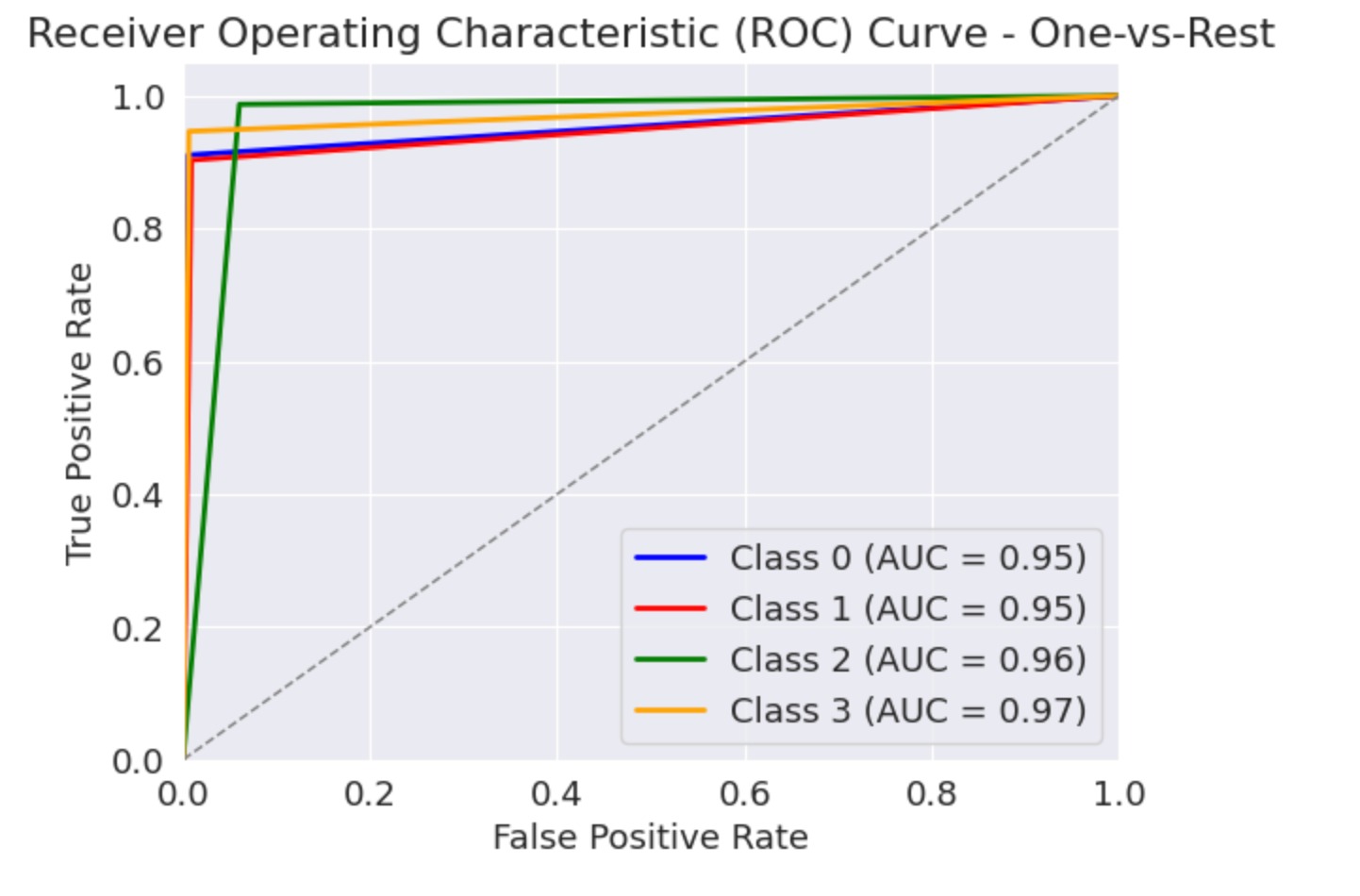}}
\caption{AUC for the multiclass dataset with histogram equalized images}
\label{fig11}
\end{figure}

The histogram equalization method is also applied tuberculosis dataset. Table IV showcases the results of the proposed approach when this method is applied as pre-processing to the input images. Fig. 12 demonstrates the AUC value for the tuberculosis dataset upon the application of histogram-equalized images. 
\begin{table}
\centering
\caption{Results of the histogram equalized images in the proposed Network on the tuberculosis dataset}
\begin{tabular}{lcccc}
\toprule
\textbf{Disease} & Recall & Precision & F1-Score & AUC \\
\midrule
Normal & 0.9967 & 0.9983 & 0.9975 & 0.99 \\
Tuberculosis & 0.9910 & 0.9821 & 0.9865 & 0.99 \\
\bottomrule
\end{tabular}
\end{table}

\begin{figure}[htbp]
\centerline{\includegraphics[width=0.5\textwidth]{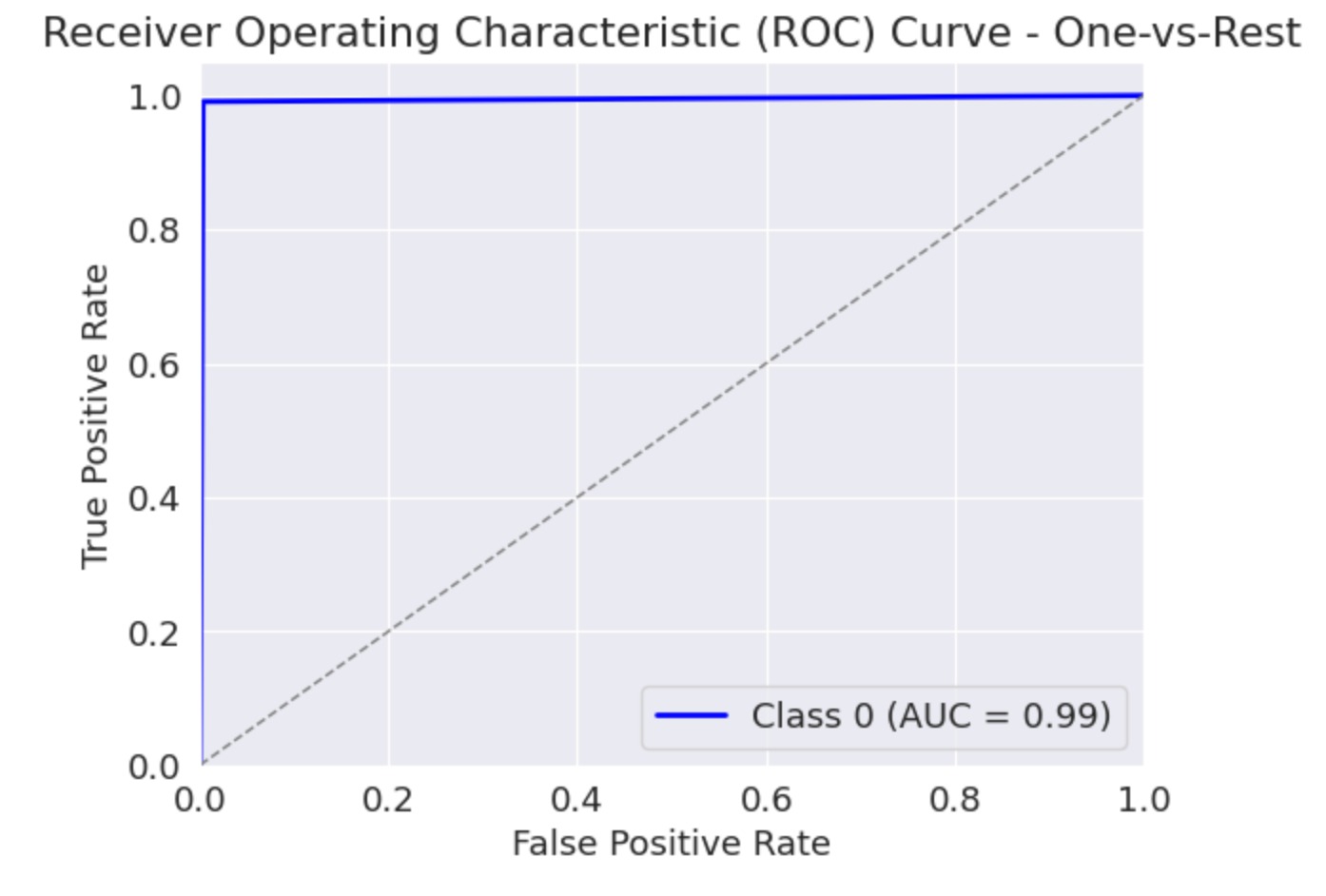}}
\caption{AUC for the tuberculosis dataset with histogram equalized images}
\label{fig12}
\end{figure}

In the tuberculosis dataset, this method surpasses state-of-the-art results. The network not only performs with state-of-the-art efficiency on the tuberculosis dataset but also demonstrates its robustness in the multiclass dataset, a tough dataset to deal with. 

The paper also experiments with a hybrid approach involving the fusion of histogram equalization and adaptive thresholding to leverage both methods in order to provide a more efficient and adaptable methodology for chest X-ray classification. The images after applying this method are seen in Fig. 13. 
\begin{figure}[htbp]
\centerline{\includegraphics[width=0.4\textwidth, height = 5cm]{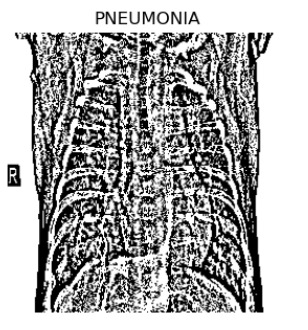}}
\caption{Input images with histogram equalization + thresholding}
\label{fig13}
\end{figure}

This method is tested on the multiclass dataset and the results are tabulated in Table V. The results obtained are not competent with state-of-the-art. The recall values for each class are much lower than the previously experimented preprocessing methods. The F1-scores and AUC values for Covid-19 and Pneumonia are poorer compared to other methods. This approach only performs competently in the tuberculosis class, which fails to provide proof of its robustness and adaptability. Fig. 14 gives the AUC values for the multiclass dataset with hybrid pre-processing techniques applied. 

\begin{table}
\centering
\caption{Results of the hybrid processed images in the proposed Network on the multiclass dataset}
\begin{tabular}{lcccc}
\toprule
\textbf{Disease} & Recall & Precision & F1-Score & AUC \\
\midrule
Covid-19 & 0.7317 & 0.9574 & 0.8295 & 0.86 \\
Normal & 0.8576 & 0.9044 & 0.8804 & 0.92 \\
Pneumonia & 0.9681 & 0.9403 & 0.9540 & 0.94 \\
Tuberculosis & 0.9933 & 0.8757 & 0.9308 & 0.99 \\
\bottomrule
\end{tabular}
\end{table}

\begin{figure}[htbp]
\centerline{\includegraphics[width=0.5\textwidth]{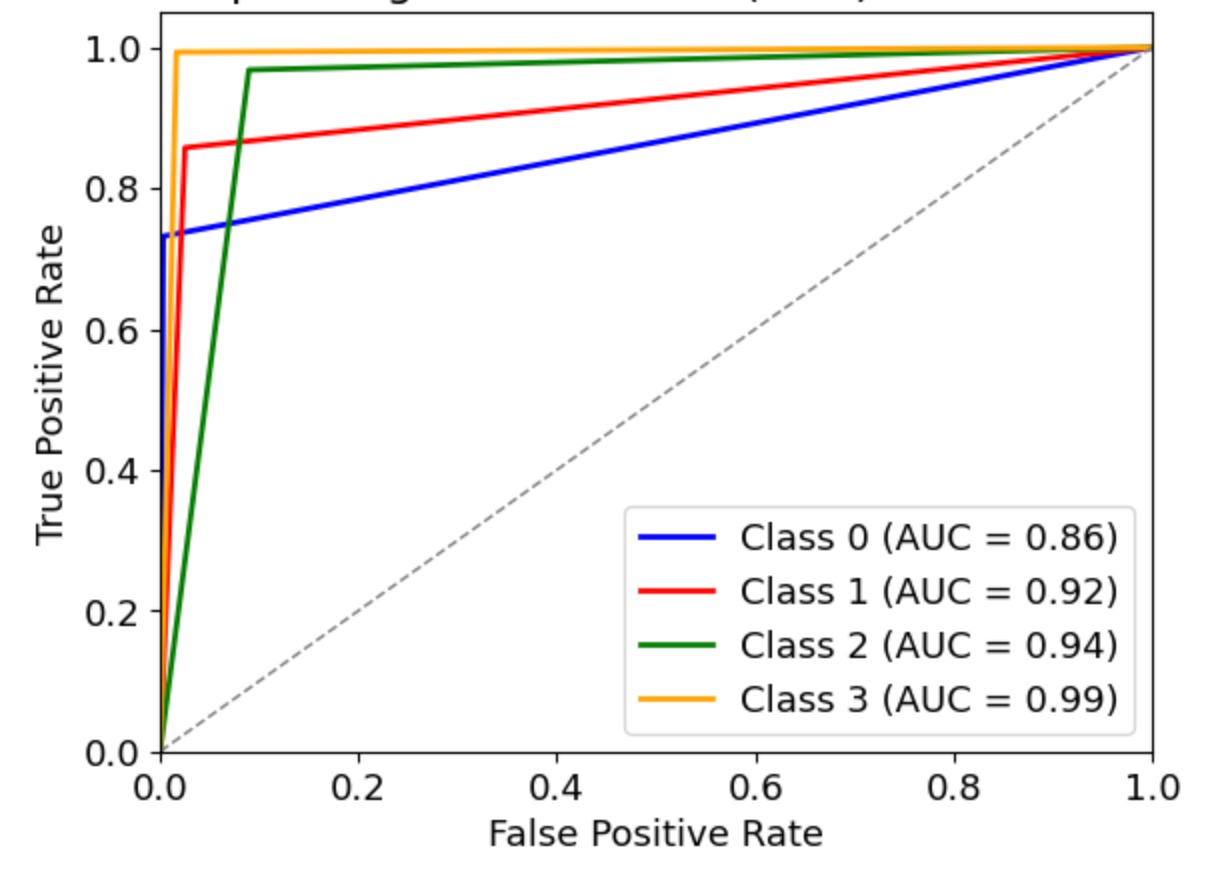}}
\caption{AUC for the multiclass dataset with hybrid images}
\label{fig14}
\end{figure}

The hybrid approach when evaluated on the tuberculosis dataset yields poor results again as compared to other methods on the tuberculosis dataset. The F1 scores (0.959) and recall(0.9286) values for the tuberculosis class are very low, compared to the competent methods. This method doesn't provide an efficient approach to tackle the problem at hand. Table VI summarises the performance metrics for the tuberculosis dataset and Fig. 15 gives the AUC values for the same. The AUC value for the tuberculosis dataset is 0.96 as evident from Fig. 15. 

\begin{table}
\centering
\caption{Results of the hybrid images in the proposed Network on the tuberculosis dataset}
\begin{tabular}{lcccc}
\toprule
\textbf{Disease} & Recall & Precision & F1-Score & AUC \\
\midrule
Normal & 0.9983 & 0.9848 & 0.9915 & 0.96 \\
Tuberculosis & 0.9286 & 0.9915 & 0.9590 & 0.96 \\
\bottomrule
\end{tabular}
\end{table}

\begin{figure}[htbp]
\centerline{\includegraphics[width=0.5\textwidth]{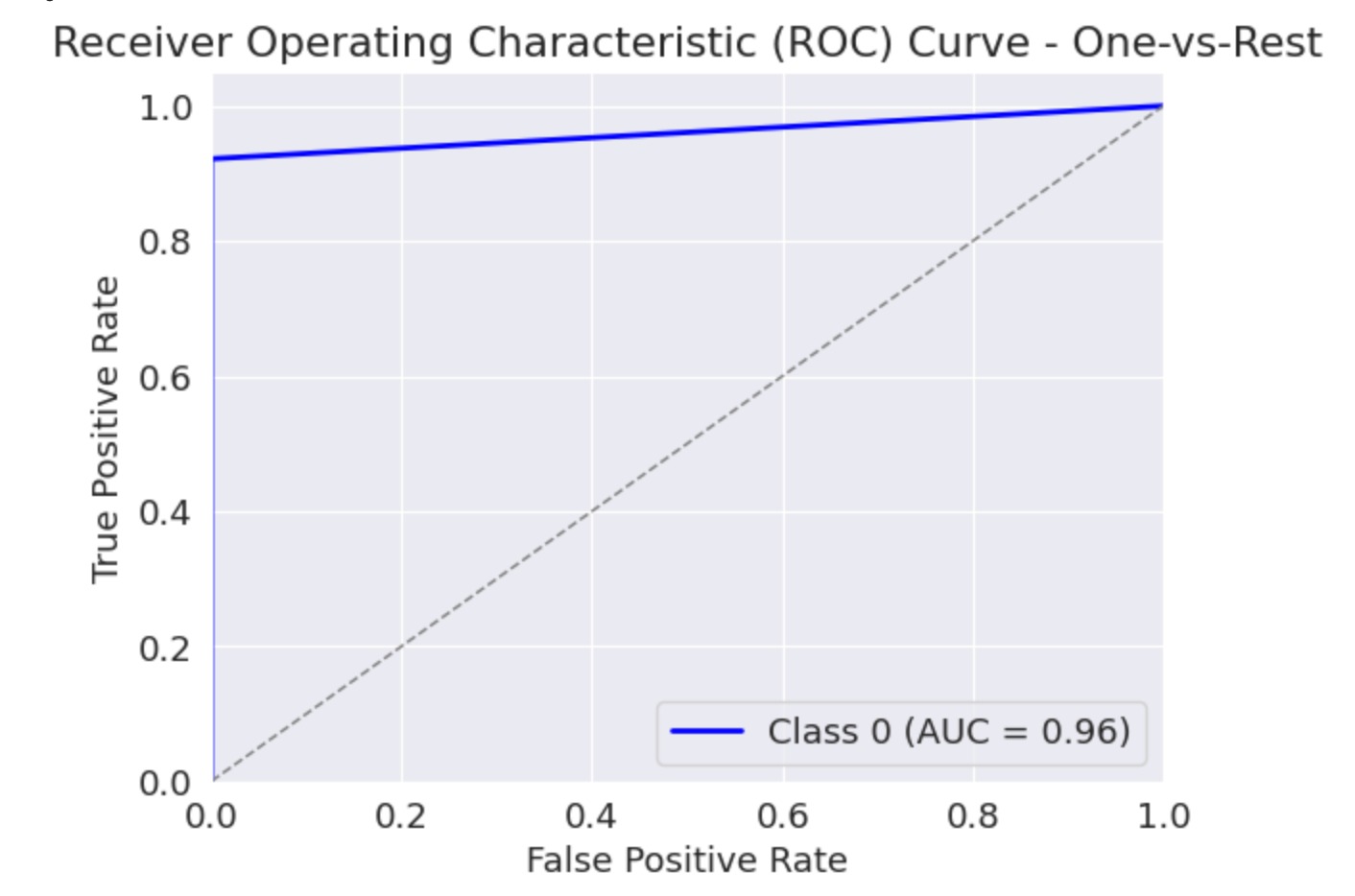}}
\caption{AUC for the tuberculosis dataset with hybrid images}
\label{fig15}
\end{figure}

Analyzing the results obtained after experimentation on the processing methods, it can be interpreted that the most efficient method for the classification task of chest X-rays, which maintains a balance between high precision, high recall and adaptability to variations in the dataset is the histogram equalization method. Hence, it will be the method used for pre-processing in this paper in the proposed network. 

The proposed approach is evaluated as follows: 
\begin{enumerate}
    \item The network is tested on the multiclass classification dataset consisting of 7135 chest X-ray images, along with a comparison with other pre-trained models. 
    \item The network is tested on the tuberculosis dataset from the Hamad Medical Corporation containing 6300 chest X-ray images. 
    \item The network is also tested on the Pneumonia dataset containing 5863 chest X-ray images. 
    \item The network is evaluated on the benchmark NIH dataset.
    \item The proposed approach is finally compared with other state-of-the-art methods to exhibit its competitiveness.  
\end{enumerate}

\begin{table}[t]
  \centering
  \caption{Comparison of Performance Metrics for the Multiclass Dataset on Different Networks}
  \label{tab:performance}
  \begin{tabular}{cccccccccc}
    \toprule
    \multirow{2}{*}{Method} & \multicolumn{3}{c}{Pneumonia} & \multicolumn{3}{c}{Tuberculosis} & \multicolumn{3}{c}{Covid-19} \\
    \cmidrule(lr){2-4} \cmidrule(lr){5-7} \cmidrule(lr){8-10}
    & Rec & Pre & F1 & Rec & Pre & F1 & Rec & Pre & F1 \\
    \midrule
    DenseNet & 0.08 & \textbf{1} & 0.14 & 0.38 & 0.75 & 0.5 & 0.1 & 0.46 & 0.16 \\
    VGG19 & \textbf{0.98} & 0.95 & 0.97 & 0.95 & \textbf{0.97} & 0.96 & 0.93 & \textbf{0.93} & 0.93 \\
    InceptionNet & \textbf{0.99} & 0.76 & 0.86 & \textbf{0.98} & 0.8 & 0.88 & 0.98 & \textbf{0.98} & 0.98 \\
    ResNet & \textbf{0.94} & 0.88 & 0.91 & \textbf{0.98} & 0.91 & 0.94 & 0.9 & \textbf{0.99} & 0.94 \\
    EfficientNet & \textbf{1} & 0.75 & 0.86 & \textbf{1} & 0.95 & 0.98 & \textbf{0.99} & 0.97 & 0.98 \\
    \textbf{Proposed CNN} & \textbf{0.98} & \textbf{0.98} & \textbf{0.97} & \textbf{0.97} & \textbf{0.97} & \textbf{0.95} & \textbf{0.96} & \textbf{0.97} & \textbf{0.92} \\
    \bottomrule
  \end{tabular}
\end{table}

From Table VII, it is clear that the proposed network performs better in all performance metrics on the multiclass classification dataset. Table VII demonstrates different methods, predominantly pre-trained models that were experimented with in this paper along with the proposed network. The table gives the results for all three classes viz. Pneumonia, Tuberculosis and Covid-19, with the precision, recall, and F1-score for all the networks that were tested. Using the proposed network, recall scores of 0.97, 0.96 and 0.89 were obtained for the three classes, as mentioned in the respective order.

As specified in Fig. 6, the lower distribution of COVID-19 chest X-ray images can be attributed to its slightly lower recall value, but in spite of that, it performs at a significantly competent level. Due to the perfect distribution of data and the data pre-processing and augmentation techniques applied, the recall values for Pneumonia and Tuberculosis are state-of-the-art. The precision scores too are higher than almost all the tested networks, for all the three diseases in the dataset. Moreover, the table also gives the F1 scores, which also signifies the superior and robust performance of the network. \par 
Table VIII showcases the results of the proposed approach on the pneumonia dataset, achieving an AUC of 0.98 with the same demonstrated in Fig. 16. 

The efficiency of the proposed approach is made evident in Table IX, with competitive results on the NIH dataset with an AUC of 0.889 as also seen in Fig. 17. 
the superiority of the proposed model is also testified by the tuberculosis dataset. Fig. 18 shows the results on the tuberculosis dataset with an AUC of 0.92, as also seen from Table X with their high recall and precision values.

\begin{table}
\centering
\caption{Results of the Proposed Network on the Pneumonia dataset}
\begin{tabular}{lcccc}
\toprule
\textbf{Disease} & Recall & Precision & F1-Score & AUC \\
\midrule
Pneumonia & 0.994 & 0.933 & 0.962 & 0.98 \\
\bottomrule
\end{tabular}
\end{table}

\begin{figure}[htbp]
\centerline{\includegraphics[width=0.45\textwidth]{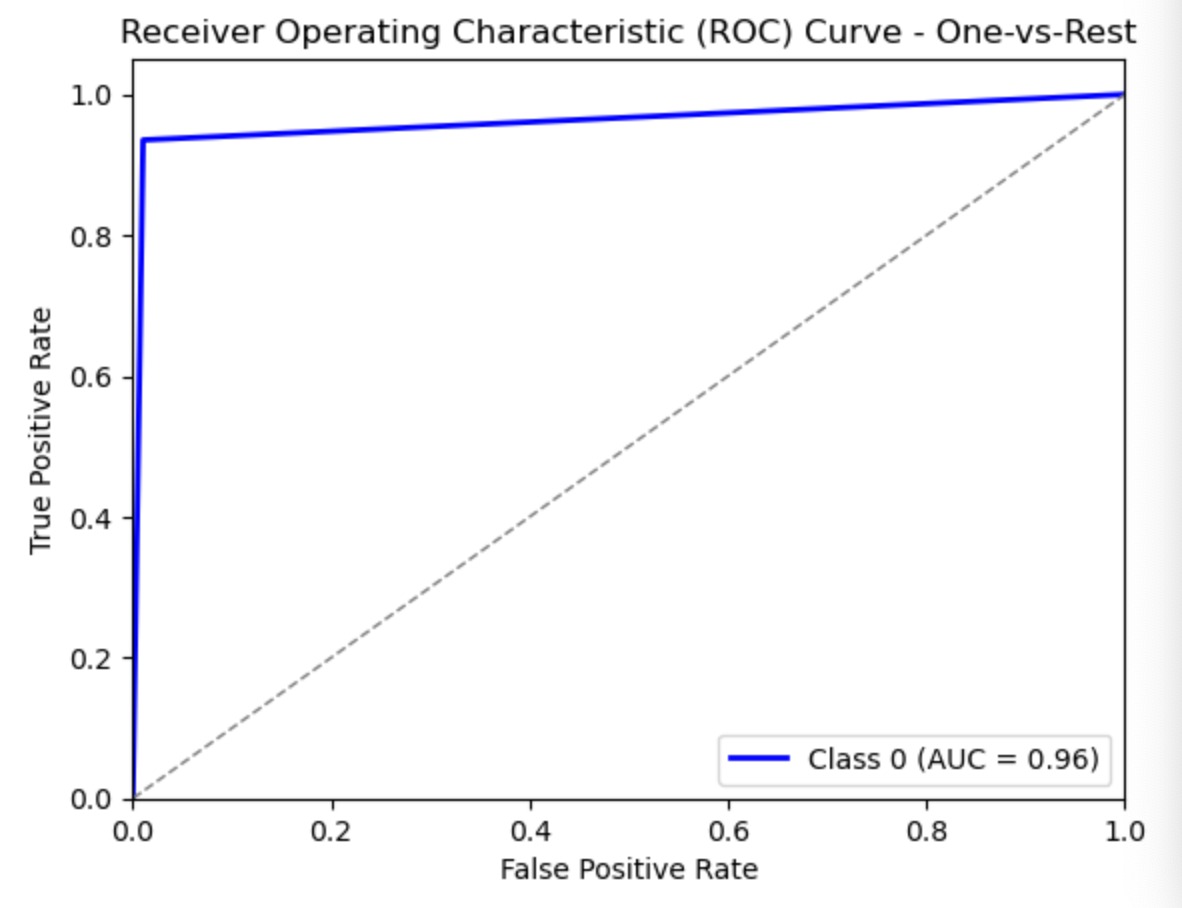}}
\caption{AUC for the pneumonia dataset}
\label{fig16}
\end{figure}

\begin{table}
\centering
\caption{Results of the Proposed Network on the benchmark NIH dataset}
\begin{tabular}{lcccc}
\toprule
\textbf{Disease} & Recall & Precision & F1-Score & AUC \\
\midrule
Pneumonia & 0.773 & 0.730 & 0.751 & 0.889 \\
\bottomrule
\end{tabular}
\end{table}

\begin{figure}[htbp]
\centerline{\includegraphics[width=0.45\textwidth]{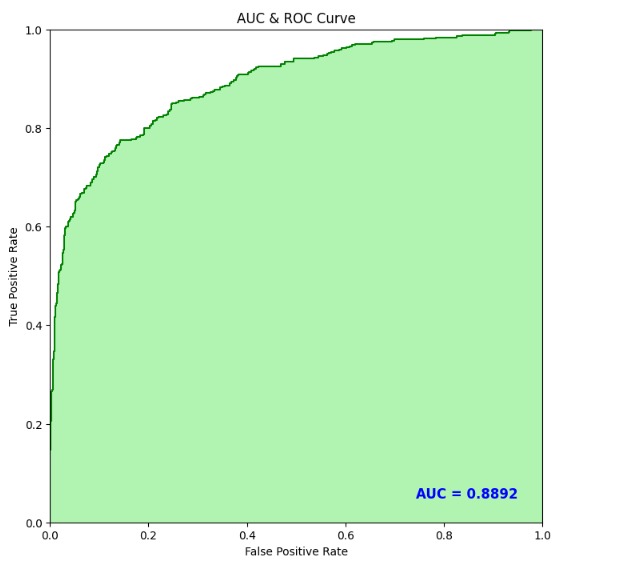}}
\caption{AUC for the NIH dataset}
\label{fig17}
\end{figure}
The paper demonstrates how the network proposed and used is efficient in multi-class classification, compared to other networks in binary as well as multi-class classification. 



\begin{table}
\centering
\caption{Results of the Proposed Network on the Tuberculosis dataset}
\begin{tabular}{lccc}
\toprule
\textbf{Disease} & Recall & Precision & F1-Score\\
\midrule
Tuberculosis & 0.9308 & 0.9098 & 0.9202 \\
\bottomrule
\end{tabular}
\end{table}

\begin{figure}[htbp]
\centerline{\includegraphics[width=0.45\textwidth]{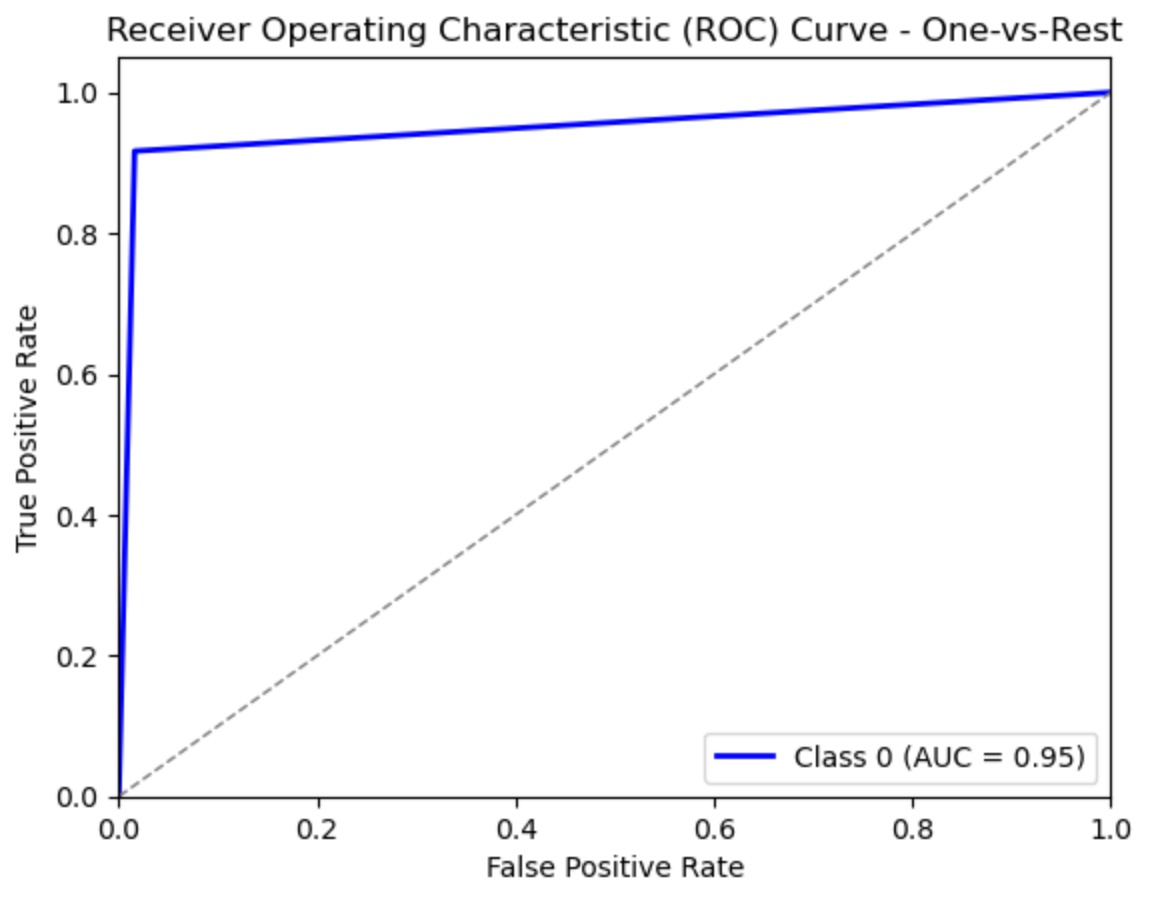}}
\caption{AUC for the tuberculosis dataset}
\label{fig19}
\end{figure}


The proposed approach is also tested on the benchmark NIH dataset. This NIH Chest X-ray Dataset is comprised of 112,120 X-ray images with disease labels from 30,805 unique patients. This dataset provides testimony to the competitiveness of the model. There are 15 different classes in this dataset. This research focuses on pneumonia, tuberculosis and COVID-19 diseases. The proposed method achieves state-of-the-art results on the NIH dataset too.

From Table XI, we obtain a detailed comparison with state-of-the-art works and research papers. We are able to surpass the results of previous state-of-the-art papers by achieving a better AUC of 0.88 with the NIH dataset as compared to Varshini et al. \cite{10}, Antin et al. \cite{9}, Rajpurkar et al. \cite{11} and Kumar et al. \cite{12} who obtained significantly lower AUC values than the previous methods. 

\begin{table*}[!h]
\centering
\caption{Comparison of Chest X-ray Models}
\begin{tabular}{cccc}
\toprule
\textbf{Previous Research} & Methodology & Dataset & Result \\
\midrule
\textbf{Varshini et al. \cite{10}} & DenseNet 169 + SVM Classifier & NIH dataset & \textbf{AUC: 0.800} \\
\textbf{Antin et al. \cite{9}} & DenseNet 121 + Logistic Regression & NIH dataset & \textbf{AUC: 0.600} \\
\textbf{Rajpurkar et al. \cite{11}} & CheXNet & NIH dataset & \textbf{AUC: 0.768} \\
\textbf{Kumar et al. \cite{12}} & Cascade Network with Binary Relevance and Pairwise Error &  NIH dataset & \textbf{AUC: 0.700, 0.637} \\
\textbf{Souid et al. \cite{24}} & Modified MobileNetV2 & NIH dataset & \textbf{0.811 AUC} \\
\textbf{Munadi et al. \cite{14}} & EfficientNetB4 with HEF & Shenzen and Montogomery Datasets & \textbf{Accuracy: 93.4\% for TB} \\
\textbf{Norval et al. \cite{15}} & Segmentation + CNN & Shenzen and Montogomery Datasets & \textbf{Accuracy: 92.54\% for TB} \\
\textbf{Ilwa et al. \cite{16}} & Ensemble learning & Shenzen and Montogomery Datasets & \textbf{Accuracy: 93\%} \\
\textbf{Gozes et al. \cite{20}} & MetaChexNet & Shenzen and Montogomery Datasets & \textbf{0.937 AUC} \\
\textbf{Basu et al. \cite{21}} & Grad-CAM & Multiclass dataset & \textbf{90.13\% accuracy }\\
\textbf{AbdElhamid et al. \cite{25}} & Three-stage methodology & Multiclass dataset & \textbf{0.99 F1-Score} \\
\textbf{Malik et al. \cite{26}} & CDC Net & Multiclass dataset & \textbf{0.980 AUC} \\
\textbf{Ibrahim et al. \cite{29}} & Classification using AlexNet & Multiclass dataset & \textbf{0.97 recall} \\
\textbf{ Khan et al. \cite{22}} & Coronet model & Pneumonia dataset & \textbf{0.93 precision} \\
\textbf{ Zebin et al. \cite{23}} & Transfer learning & Covid-19 dataset & \textbf{96\% accuracy} \\
\textbf{Li et al. \cite{19}} & MHA-CoroCapsule & Covid-19 dataset & \textbf{0.97 recall} \\
\textbf{Batista et al. \cite{17}} & SVM, Random forest, LR & Custom database & \textbf{AUC: 0.847} \\
\textbf{Kant et al. \cite{13}} & Neural Network Cascade method & Microscopic images of TB cells & \textbf{Recall: 0.83}, Precision: 0.68 \\

\bottomrule
\end{tabular}
\end{table*}

Li et al. \cite{19} propose MHA-CoroCapsule, a unique AI model that diagnoses COVID-19 from chest X-ray pictures by applying a multi-head attention routing approach. Their model achieved 97.28\% accuracy, 0.97 recall, and 0.97 precision. 

Gozes et al. \cite{20} also carried out significant research in their work, where they presented a model named MetaChexNet. They trained a DenseNet-121 CNN on the NIH dataset. Their results show the effectiveness of transfer learning on small datasets for tuberculosis and highlight the value of combining picture data with information for more precise illness classification.

The study conducted by Basu et al. \cite{21} used a deep convolutional neural network that had been trained beforehand to categorize chest X-rays into four groups: normal, pneumonia, various illnesses, and COVID-19. A 5-fold cross-validation showed promise, producing a COVID-19 detection accuracy of 90.13\% + 0.14 as a whole. Furthermore, Gradient Class Activation Map (Grad-CAM) implementation showed strong agreement with clinical findings, which was supported by expert validation. 

 Khan et al. \cite{22} used the CoroNet model, a Deep Convolutional Neural Network based on the Xception architecture from the ImageNet dataset that has already been trained. It was trained using a collection of chest X-ray pictures from databases that were accessible to the general public, including cases with COVID-19 and other pneumonia-related illnesses.CoroNet performed admirably, as seen by its high accuracy score of 89.6\%. A notable result for the 4-class classification task (COVID vs. Pneumonia bacterial vs. pneumonia viral vs. normal) was that the model showed a precision rate of 0.93 and a recall rate of 0.98. CoroNet achieved a 95\% accuracy rate for the three classes of classification (COVID, Pneumonia, and Normal).

In the research performed by Zebin et al. \cite{23}, three different pre-trained convolutional backbones—VGG16, ResNet50, and EfficientNetB0 - as part of a transfer learning technique to categorize COVID-19 chest X-ray pictures were used. They used labeled data from numerous sources to conduct their research on publicly accessible datasets from chest X-rays. The classification accuracies of 90\%, 94.3\%, and 96.8\% for VGG16, ResNet50, and EfficientNetB0, respectively, showed good results. 

Souid et al. \cite{24} used a modified MobileNet V2 model for the categorization and prediction of lung diseases in frontal thoracic X-rays. They used the NIH Chest-Xray-14 database for training and evaluation, utilizing transfer learning and metadata. With an average AUC of 0.811 and an accuracy higher than 90\%, their method produced significant results.

AbdElhamid et al. \cite{25} demonstrated their methodology in three phases. The three main stages of the methodology were initial noise removal, image scaling, feature extraction using a pre-trained Xception model with global average pooling and activation layers, and classification using a softmax layer. The analysis made use of the multiclass dataset made up of 7395 photos divided into three categories: pneumonia, normal, and COVID-19. The test had a 99.3\% accuracy rate, a 0.99 sensitivity rate, a 0.99 specificity rate, and a 0.99 F1-Score, which are all impressive and state-of-the-art results. 

The work experimented by Malik et al. \cite{26} used chest x-ray pictures to construct and evaluate a deep learning model dubbed CDC Net for the categorization of five chest illnesses, including COVID-19, lung cancer (LC), pneumothorax, tuberculosis (TB), and pneumonia. In order to correctly detect these disorders, the methodology used a CNN including residual network ideas and dilated convolution. With CDC Net achieving an AUC of 0.9953, an accuracy of 99.39\%, a recall of 0.98, and a precision of 0.99, the results showed exceptional performance.

An impressive novel approach was proposed by \cite{27}. Two methods were used in this study's methodology: (i) employing standard chest radiographs; and (ii) adding a threshold filter to those same radiographs. The SoftMax classifier was initially used with traditional DLS models including AlexNet, VGG16, VGG19, and ResNet50, with VGG19 producing the greatest classification accuracy at 86.97\%. Then, a tailored VGG19 network was suggested using an Ensemble Feature Scheme (EFS), which incorporated deep features gained from transfer learning with hand-crafted features from CWT, DWT, and GLCM. The best accuracy was attained by VGG19 and Random Forest, which had a 95.70\% accuracy rate for benchmark chest radiographs and a remarkable 97.94\% accuracy rate for threshold-filtered chest radiographs. 

Ibrahim et al. \cite{29} approached the classification task using the categorization of chest X-ray (CXR) pictures associated to COVID-19, non-COVID-19 viral pneumonia, bacterial pneumonia, and normal CXR scans was done using a pretrained AlexNet model. The collection was made up of CXR images that were taken from several open databases. The model was trained on a variety of classification tasks, including scenarios with two, three, and four possible outcomes. Notably, the results showed outstanding sensitivity and accuracy, with the model differentiating COVID-19 pneumonia from regular CXR scans with 99.16\% accuracy, 0.97 sensitivity, and 1 specificity. These results show how deep learning techniques may be used to quickly and accurately screen for respiratory illnesses, which is a critical necessity during the COVID-19 epidemic. 

A approach including using architectures, such as VGG-16, to identify between healthy individuals, pneumonia patients, and tuberculosis cases, proposed by  Shelke et al. \cite{30}, was performed with a test accuracy of an astonishing 95.9\%. To further categorize COVID-19 cases, they used DenseNet-161, reaching a phenomenal test accuracy of 98.9\%. Additionally, ResNet-18 demonstrated its usefulness in determining the severity of COVID-19, attaining test accuracy of up to 76\%. 

\section{Conclusion}

In this paper, we employ modern deep learning techniques, enhanced by data augmentation and advanced image processing methods, to introduce a CNN model for efficient chest disease detection from X-ray images. The aim of this research was to introduce a novel and robust network that would work in all types of use cases and reduce the number of false positives and negatives, leading to accurate diagnostic precision. Our experiments reveal highly competitive results in terms of performance metrics, surpassing previous approaches. Incorporating image enhancement techniques enhances the model's image-learning capabilities. The research demonstrates the effectiveness of our multi-class classification network in segregating chest X-ray scans and have rigorously tested it across multiple datasets, consistently achieving state-of-the-art results. The histogram equalization pre-processing performs the best on all datasets, achieving robust and efficient classification performances on all datasets. This paper achieves state-of-the-art results on the benchmark NIH data. The proposed network obtains an AUC of 0.98 on the multiclass dataset, and 0.99 AUC on the tuberculosis dataset, 0.98 AUC on pneumonia dataset and 0.889 AUC on the benchmark NIH dataset. Moreover, compared to traditional manual detection methods, our automated approach offers advantages such as improved accuracy, efficiency, and reduced reliance on subjective human interpretation. Future work will explore additional image enhancement techniques and engage medical experts for comprehensive evaluation. 


\printbibliography

\end{document}